\documentclass[12pt]{article}
\usepackage{amsmath}
\usepackage{amssymb}
\usepackage{epsfig}
\begin{document}

\numberwithin{equation}{section}

\begin{table}[t]
  \begin{flushright}
     CBPF-NF-008/07    \\
  \end{flushright}
\end{table}

\title{Chern-Simons $ AdS_{5} $ supergravity in a Randall-Sundrum 
background}

\author{Richard S. Garavuso\thanks{garavuso@cbpf.br} \, and 
Francesco Toppan\thanks{toppan@cbpf.br} \\
\emph{\normalsize Centro Brasileiro de Pesquisas F\'{\i}sicas}
\\
\emph{\normalsize Rio de Janeiro, RJ 22290-180, Brazil}}

\date{}

\maketitle
\thispagestyle{empty}

\begin{abstract}

Chern-Simons AdS supergravity theories are gauge theories for the 
super-AdS group.
These theories possess a fermionic symmetry which differs from standard 
supersymmetry.
In this paper, we study five-dimensional Chern-Simons AdS supergravity in 
a Randall-Sundrum scenario with two Minkowski 3-branes.
After making modifications to the $ D = 5 $ Chern-Simons AdS supergravity 
action and fermionic symmetry transformations, we obtain a 
$ \mathbb{Z}_{2} $-invariant total action 
$ S = \tilde{S}_{bulk} + S_{brane} $ and fermionic transformations
$ \tilde{\delta}_{\epsilon} $.
While $ \tilde{\delta}_{\epsilon} \tilde{S}_{bulk} = 0 $,
the fermionic symmetry is broken by $ S_{brane} $.
Our total action reduces to the original Randall-Sundrum model when
$ \tilde{S}_{bulk} $ is restricted to its gravitational sector.
We solve the Killing spinor equations for a bosonic configuration with 
vanishing $ su(N) $ and $ u(1) $ gauge fields.

\end{abstract}

\pagebreak

\section{Introduction}

\emph{Chern-Simons AdS supergravity} 
\cite{CSADS5,CSAdSseveneleven,CSAdSodd}
theories can be constructed only in odd spacetime dimensions.
As the name implies, they are gauge theories for supersymmetric 
extensions of the AdS group.\footnote
{
The AdS group in dimension $ D \geq 2 $ is $ SO(D-1,2) $.  
The corresponding super-AdS groups are given in \cite{CSAdSodd}.
For $ D \! = \! 5 $ and $ D \! = \! 11 $, the super-AdS groups are 
respectively $ SU(2,2|N) $ and $ OSp(32|N) $.
}
They have a fiber bundle structure and hence are potentially
renormalizable \cite{CSAdSseveneleven}.
The dynamical fields form a single adS superalgebra-valued connection and 
hence the supersymmetry algebra closes automatically \emph{off-shell} 
without requiring auxiliary fields \cite{off-shell}.
The Lagrangian in dimension $ D = 2n - 1 $ is a Chern-Simons 
$ (2n-1) $-form for the super-adS connection and is a polynomial of order 
$ n $ in the corresponding curvature.
Unlike standard supergravity theories, there can be a mismatch
between the number of bosonic and fermionic degrees of freedom.\footnote
{
For example, in 
$ D \! = \! 5 $ \emph{Chern-Simons AdS supergravity} \cite{CSADS5}, 
the number of bosonic degrees of freedom $ (N^{2} + 15) $ is equal to 
the number of fermionic degrees of freedom $ (8N) $ 
only for $ N \! = \! 3 $ and $ N \! = \! 5 $.
}
For this reason, the `supersymmetry' of Chern-Simons AdS supergravity 
theories is perhaps better referred to as a fermionic symmetry.

$ D \! = \! 11 $, $ N \! = \! 1 $ 
\emph{Chern-Simons AdS supergravity}
may correspond to an off-shell supergravity limit of M-theory 
\cite{CSAdSseveneleven, CSAdSodd}.
It has expected features of M-theory which are not shared by 
$ D \! = \! 11 $ \emph{Cremmer-Julia-Scherk (CJS) supergravity} 
\cite{CJS}.
These features include an $ osp(32|1) $ superalgebra \cite{Town:p-Brane}
and higher powers of curvature \cite{curvature}.
\emph{Ho\v{r}ava-Witten theory} \cite{HorWit} 
is obtained from CJS supergravity by compactifying on an 
$ S^{1} / \mathbb{Z}_{2} $ orbifold and requiring gauge and gravitational 
anomalies to cancel. 
This theory gives the low energy, strongly coupled limit of the heterotic 
$ E_{8} \times E_{8} $ string theory.
In light of the above discussion, it would be interesting to reformulate 
Ho\v{r}ava-Witten theory with 
$ D \!= \!11 $, $ N \! = 1 \! $ Chern-Simons AdS supergravity.

Reformulating Ho\v{r}ava-Witten theory as described above may prove to be 
difficult.
It is simpler to compactify the five-dimensional version 
of Chern-Simons AdS supergravity on an 
$ S^{1} / \mathbb{Z}_{2} $ orbifold and ignore anomaly cancellation 
issues.
Canonical sectors of  
$ D \! = \! 5 $ Chern-Simons AdS supergravity
have been investigated in locally 
$ AdS_{5} $ backgrounds possessing a spatial boundary with topology 
$ S^{1} \times S^{1} \times S^{1} $ located at infinity 
\cite{CSADS5_OneBraneRand}.
In this paper, as a preamble to reformulating Ho\v{r}ava-Witten theory,  
we will study 
$ D \! = \! 5 $ Chern-Simons AdS supergravity 
in a Randall-Sundrum background with two Minkowski 3-branes 
\cite{RandSund:ALarge}.  
We choose coordinates
$ x^{\mu} = \left( x^{\bar{\mu}}, x^{5} \right) $
to parameterize the five-dimensional spacetime manifold.\footnote
{
We use indices
$ \mu, \nu, \ldots = 0,1,2,3,5 $ for local spacetime and
$ a,b, \ldots = \dot{0}, \dot{1}, \dot{2}, \dot{3}, \dot{5} $
for tangent spacetime.
The corresponding metrics,
$ g_{\mu \nu} $ and
$ \eta_{ab} = \textrm{diag} (-1,1,1,1,1)_{ab} $,
are related by
$ g_{\mu \nu} = e_{\mu}{}^{a} e_{\nu}{}^{b} \eta_{ab} $, where
$ e_{\mu}{}^{a} $ is the f\"{u}nfbein.
Barred indices
$ \bar{\mu}, \bar{\nu}, \ldots = 0,1,2,3, $ and
$ \bar{a}, \bar{b}, \ldots = \dot{0}, \dot{1}, \dot{2}, \dot{3} $
denote the four-dimensional counterparts of
$ \mu, \nu, \ldots $ and
$ a,b, \ldots $, respectively.
}
In terms of these coordinates, the background metric takes the form
\begin{equation}
\label{metric}
g_{\mu \nu} dx^{\mu} dx^{\nu}
  =  \mathfrak{a}^{2}(x^{5}) \eta^{(4)}_{ \bar{\mu} \bar{\nu} }
      dx^{ \bar{\mu} } dx^{ \bar{\nu} }
    + (dx^{5})^{2},
\end{equation}
where
$ \eta^{(4)}_{ \bar{\mu} \bar{\nu} }
  = \rm{diag}(-1,1,1,1)_{ \bar{\mu} \bar{\nu} } $,
$ \mathfrak{a}(x^{5}) \equiv \exp( -|x^{5}| / \ell ) $
is the \emph{warp factor}, and
$ \ell $ is the $ AdS_{5} $ curvature radius.
The coordinate $ x^{5} $ parameterizes an 
$ S^{1} / \mathbb{Z}_{2} $ orbifold, where the circle 
$ S^{1} $ has radius $ \rho $ and 
$ \mathbb{Z}_{2} $ acts as 
$ x^{5} \rightarrow - x^{5} $.
We choose the range 
$ - \pi \rho \leq x^{5} \leq \pi \rho $ 
with the endpoints identified as 
$ x^{5} \simeq x^{5} + 2 \pi \rho $.
The Minkowski 3-branes are located at the 
$ \mathbb{Z}_{2} $ fixed points 
$ x^{5} = 0 $ and $ x^{5} = \pi \rho $.
These 3-branes have corresponding tensions 
$ \mathcal{T}^{(0)} $ and $ \mathcal{T}^{(\pi \rho)} $
and may support $ (3+1) $-dimensional field theories.

This paper is organized as follows:  
In Section \ref{Bulk}, we construct a 
$ \mathbb{Z}_{2} $-invariant bulk theory.
This bulk theory is obtained by making modifications to the
$ D \! = \! 5 $ Chern-Simons AdS supergravity action and fermionic 
symmetry transformations which allow consistent orbifold conditions to be
imposed. 
The variation of the resulting bulk action $ S_{bulk} $ under the 
resulting fermionic transformations $ \delta_{\epsilon} $ vanishes 
everywhere except at the $ \mathbb{Z}_{2} $ fixed points.
We calculate $ \delta_{\epsilon} S_{bulk} $ in Section \ref{Calculation}.
In Section \ref{ModifiedBulk}, we modify 
$ S_{bulk} $ and $ \delta_{\epsilon} $
to obtain a modified $ \mathbb{Z}_{2} $-invariant bulk theory.
The modified bulk action $ \tilde{S}_{bulk} $ 
is invariant under the modified fermionic transformations 
$ \tilde{\delta}_{\epsilon} $.
In Section \ref{Brane}, we 
complete our model by adding the brane action $ S_{brane} $.
We show in Section \ref{Connection} that our total action
\begin{equation}
S = \tilde{S}_{bulk} + S_{brane}
\end{equation}
reduces to the original Randall-Sundrum model \cite{RandSund:ALarge}
when $ \tilde{S}_{bulk} $ is restricted to its gravitational sector. 
In Section \ref{Killing}, we solve the Killing spinor equations for 
a purely bosonic configuration with vanishing $ su(N) $ and $ u(1) $ 
gauge fields.
Our concluding remarks are given in Section \ref{Conclusion}.
Finally, in the Appendix, we work out the f\"{u}nfbein, spin connection, 
curvature 2-form components, Ricci tensor, and Ricci scalar for our 
metric (\ref{metric}).

\section{\label{Bulk}$ \mathbb{Z}_{2} $-invariant bulk theory}

In this section, we construct a $ \mathbb{Z}_{2} $-invariant bulk theory.
The bulk theory is obtained by making modifications to the 
$ D \! = \! 5 $ Chern-Simons AdS supergravity \cite{CSADS5} 
action and fermionic symmetry transformations which allow consistent 
orbifold conditions to be imposed. 

The field content of 
$ D \! = \! 5 $ Chern-Simons AdS supergravity is
the f\"{u}nfbein 
$ e_{\mu}{}^{a} $,
the spin connection 
$ \omega_{\mu}{}^{ab} $,
the $ su(N) $ gauge connection 
$ A_{\mu} = A^{i}_{\mu} \tau_{i} $,
the $ u(1) $ gauge connection $ B_{\mu} $,
and $ N $ complex gravitini 
$ \psi_{\mu r} $
which transform as Dirac spinors in a vector representation of 
$ su(N) $.\footnote
{
We use indices $ i,j,\ldots = 1,\ldots,N^{2}-1 $ to label the 
$ N \times N $-dimensional $ su(N) $ generators $ \tau_{i} $.
The indices $ r,s,\ldots = 1,\ldots,N $ label vector representations of 
$ su(N) $.
We will use the notation 
$ A^{r}_{s} \equiv A^{i} (\tau_{i})^{r}_{s} $. 
Spinor indices $ \alpha, \beta, \ldots $ will sometimes be suppressed.
}
These fields form a connection for the adS superalgebra $ su(2,2|N) $.
The action and fermionic symmetry transformations are given in
\cite{CSADS5_OneBraneRand} in terms of the 
$ AdS_{5} $ curvature radius $ \ell $.
The only free parameter in the action is a dimensionless constant $ k $.
To allow consistent $ \mathbb{Z}_{2} $ orbifold conditions to be imposed, 
we make the following modifications:
\begin{enumerate}

\item{Rescale the $ su(N) $ and $ u(1) $ gauge connections:
\begin{equation*}
A \rightarrow g_{A} A,
\quad
B \rightarrow g_{B} B.
\end{equation*}}

\item{Replace $ g_{A} $, $ g_{B} $, $ \ell^{-1} $, and $ k $ by the 
$ \mathbb{Z}_{2} $-odd expressions\footnote
{
The \emph{signum function} $ \mathrm{sgn}(x^{5}) $ is
$ +1 $ for
$ 0 < x^{5} < \pi \rho $ and
$ -1 $ for
$ - \pi \rho < x^{5} < 0 $.
It obeys $ \mathrm{sgn}^{2}(x^{5}) = 1 $ and 
$ \partial_{5} \, \mathrm{sgn}(x^{5})
   = 2 [ \delta (x^{5}) - \delta (x^{5} - \pi \rho) ]. $
}
\begin{gather*}
G_{A} \equiv g_{A} \, \mathrm{sgn}(x^{5}),
\quad
G_{B} \equiv g_{B} \, \mathrm{sgn}(x^{5}),
\quad
L^{-1} \equiv \ell^{-1} \, \mathrm{sgn}(x^{5}),
\\
K \equiv k \, \mathrm{sgn}(x^{5}).
\end{gather*}}

\end{enumerate}
In this manner, we obtain the bulk action
\begin{equation}
S_{bulk} = S_{grav} + S_{su(N)} + S_{u(1)} + S_{ferm}, 
\end{equation}
where
\begin{align}
\nonumber
S_{grav} 
  &= {\textstyle 
     \int \frac{1}{8} K \varepsilon_{abcde}
     \left( 
          \frac{1}{ L } R^{ab} R^{cd} e^{e} 
        + \frac{2}{ 3 L^{3} } 
          R^{ab} e^{c} e^{d} e^{e} 
        + \frac{1}{ 5 L^{5} } e^{a} e^{b} e^{c} e^{d} e^{e}
     \right),
     }
\\
\nonumber
S_{su(N)} 
  &= {\textstyle 
     \int iK \ {\rm{str}}
     \left(
          G^{3}_{A} A F^{2}
        - \frac{1}{2} G^{4}_{A} A^{3} F
        + \frac{1}{10} G^{5}_{A} A^{5}
     \right),
     }
\\
\nonumber
S_{u(1)} 
  &= {\textstyle 
     \int K
     \left[
        - \left(
               \frac{1}{ 4^{2} } - \frac{1}{ N^{2} } 
          \right)
          G^{3}_{B} 
          B (dB)^{2}
        + \frac{3}{ 4 L^{2} } 
          \bigl(
                T^{a} T_{a} 
              - \frac{ L^{2} }{2} R^{ab} R_{ab}
          \bigr. 
     \right.
     }
\\
\nonumber
  &\quad \  
     {\textstyle 
     \Bigl.
          \bigl.
             - R^{ab} e_{a} e_{b}
          \bigr)
          G_{B} B
        - \frac{3}{N} G^{2}_{A} G_{B}
          F^{r}_{s} F^{s}_{r} B
     \Bigr],
     }
\\
\label{S_bulkSectors}
S_{ferm} 
   &=  {\textstyle
       \int \frac{3}{2i} K 
       \left(
            \bar{\psi}^{r}_{\alpha} \mathcal{R}^{\alpha}_{\beta}
            \nabla \psi^{\beta}_{r}
          + \bar{\psi}^{s}_{\alpha} \mathcal{F}^{r}_{s} 
            \nabla \psi^{\alpha}_{r}
       \right)
     + c.c.
       },
\end{align}
and the transformations
\begin{align}
\nonumber
\delta_{\epsilon} e^{a} 
  &= - {\textstyle \frac{1}{2} } 
       ( 
          \bar{\psi}^{r} \Gamma^{a} \epsilon_{r} 
        - \bar{\epsilon}^{r} \Gamma^{a} \psi_{r}
       ), &
\delta_{\epsilon} \omega^{ab} 
  &=  {\textstyle \frac{1}{4} }
       (
          \bar{\psi}^{r} \Gamma^{ab} \epsilon_{r}
        - \bar{\epsilon}^{r} \Gamma^{ab} \psi_{r}
       ),
\\
\nonumber
\delta_{\epsilon} \psi_{r}
  &= - \nabla \epsilon_{r}, &
\delta_{\epsilon} \bar{\psi}^{r}
  &= - \nabla \bar{\epsilon}^{r},
\\
\label{transformations}
\delta_{\epsilon} A^{r}_{s}
  &= i \left(
            \bar{\psi}^{r} \epsilon_{s}
          - \bar{\epsilon}^{r} \psi_{s}
       \right), &
\delta_{\epsilon} B
  &= i \left(
            \bar{\psi}^{r} \epsilon_{r}
          - \bar{\epsilon}^{r} \psi_{r}
       \right).
\end{align}
In these expressions,
$ \Gamma^{a} $ are the Dirac matrices\footnote
{
We choose a chiral basis for the $ 4 \times 4 $ Dirac matrices
\begin{equation*} 
\Gamma^{a}
  = \left( \Gamma^{ \bar{a} }, \Gamma^{ \dot{5} } \right)
  = \left(
           \begin{bmatrix}
              \mathbf{0}                  &  -i\sigma^{ \bar{a} }   \\
             -i \bar{\sigma}^{ \bar{a} }  &   \mathbf{0}
           \end{bmatrix},
           \begin{bmatrix}
             -\mathbf{1}             &   \mathbf{0}      \\
              \mathbf{0}             &   \mathbf{1}
           \end{bmatrix}
    \right),
\end{equation*}
where
$ \sigma^{ \bar{a} } = (\mathbf{1}, \vec{\sigma} ) $
and
$ \bar{\sigma}^{ \bar{a} } = (\mathbf{1}, - \vec{\sigma} ) $.
These matrices satisfy
$ \mathrm{tr}  
  \left(
       \Gamma_{a} \Gamma_{b} \Gamma_{c} \Gamma_{d} \Gamma_{e}
  \right)
   = -4 i \varepsilon_{abcde} $,
where 
$ \varepsilon_{abcde} $ is the Levi-Civita tensor and 
$ \varepsilon^{ \dot{0} \dot{1} \dot{2} \dot{3} \dot{5} } = 1 $. 
},
$ \Gamma^{ab}
   \equiv \frac{1}{2} 
          \left(
               \Gamma^{a} \Gamma^{b}
             - \Gamma^{b} \Gamma^{a}
          \right) $,
$ R^{ab} = d \omega^{ab} + \omega^{ac} \omega_{c}{}^{b} $
is the curvature 2-form,
$ T^{a} = de^{a} + \omega^{a}{}_{b} e^{b} $
is the torsion 2-form,
$ F = dA + G_{A} A^{2} = F^{i} \tau_{i} $
is the $ su(N) $ curvature, 
\begin{align}
\nonumber
\mathcal{R}^{\alpha}_{\beta}
 &\equiv {\textstyle
         \frac{1}{2L} T^{a} ( \Gamma_{a} )^{\alpha}_{\beta}
       + \frac{1}{4}
         ( R^{ab} + \frac{1}{ L^{2} } e^{a} e^{b} )
         ( \Gamma_{ab} )^{\alpha}_{\beta}
       + \frac{i}{4} G_{B} \, dB \, \delta^{\alpha}_{\beta}
       - \frac{1}{2} \psi^{\alpha}_{s} \bar{\psi}^{s}_{\beta}
         },
\\
\mathcal{F}^{r}_{s} 
 &\equiv {\textstyle
         F^{r}_{s} 
       + \frac{i}{N} G_{B} \, dB \, \delta^{r}_{s} 
       - \frac{1}{2} \bar{\psi}^{r}_{\beta} \psi^{\beta}_{s} 
         },
\end{align}
$ \mathrm{str} $ is a symmetrized trace satisfying
$ \mathrm{str}(\tau_{i} \tau_{j} \tau_{k}) 
  \equiv \frac{1}{2i} \mathrm{tr} 
         \left( \{ \tau_{i}, \tau_{j} \} \tau_{k} \right) $,  
$ \nabla $ is the
$ adS_{5} \times su(N) \times u(1) $ 
covariant derivative, and
\begin{align}
\nonumber
\nabla \psi_{r} 
  &\equiv {\textstyle
          \left( 
                 d + \frac{1}{4} \omega^{ab} \Gamma_{ab}
               + \frac{1}{2 L} e^{a} \Gamma_{a}
            \right)
            \psi_{r}
          - G_{A} A^{s}_{r} \psi_{s} 
          + i 
            \left(
                 \frac{1}{4} - \frac{1}{N}
            \right)
            G_{B} B \psi_{r},
          }
\\
\nabla \epsilon_{r}
  &\equiv {\textstyle
          \left(
                 d + \frac{1}{4} \omega^{ab} \Gamma_{ab}
               + \frac{1}{2 L} e^{a} \Gamma_{a}
            \right)
            \epsilon_{r}
          - G_{A} A^{s}_{r} \epsilon_{s}
          + i
            \left(
                 \frac{1}{4} - \frac{1}{N}
            \right)  
            G_{B} B \epsilon_{r}.
          }
\end{align}
Note that the results in the Appendix can be used to show that the torsion 
vanishes for our metric.

We impose the following orbifold conditions:
\begin{enumerate}

\item
{
\emph{Periodicity on $ S^{1} $.}
The fields and the fermionic parameters $ \epsilon_{r} $, denoted 
generically by $ \phi $, are required to be periodic on the circle $ S^{1} $.
That is,
\begin{equation}
\phi(x^{ \bar{\mu} }, x^{5}) = \phi(x^{ \bar{\mu} }, x^{5} + 2 \pi \rho).
\end{equation}    
}

\item
{
\emph{$ \mathbb{Z}_{2} $ parity assignments.}
The bosonic field components
\begin{equation*}
\Phi = e_{\mu}{}^{ \bar{a} }, 
        e_{5}{}^{ \dot{5} }, 
        A^{i}_{5}, 
        B_{5},
\quad
\Theta = e_{ \bar{\mu} }{}^{ \dot{5} }, 
          e_{5}{}^{ \bar{a} }, 
          A^{i}_{ \bar{\mu} }, 
          B_{ \bar{\mu} }
\end{equation*}
are chosen to satisfy
\begin{equation}
\Phi( x^{\mu}, x^{5} ) = + \Phi( x^{\mu}, -x^{5} ),
\quad
\Theta( x^{\mu}, x^{5} ) = - \Theta( x^{\mu}, -x^{5} ).
\end{equation}
That is, the  $ \Phi $ components are $ \mathbb{Z}_{2} $-even 
and the $ \Theta $ components are $ \mathbb{Z}_{2} $-odd.
For the gravitini, we require
\begin{align}
\nonumber
\label{GravitinoParity}
\Gamma^{ \dot{5} } \, \psi_{ \bar{\mu} r } ( x^{ \bar{\mu} }, x^{5} )
    &= + \, \psi_{ \bar{\mu} r } ( x^{ \bar{\mu} }, - x^{5} ),
\\
\Gamma^{ \dot{5} } \, \psi_{5 r} ( x^{ \bar{\mu} }, x^{5} )
    &= - \, \psi_{5r} ( x^{ \bar{\mu} }, - x^{5} ).
\end{align}
Finally, the fermionic parameters $ \epsilon_{r} $ are required to satisfy
\begin{equation}
\label{FermionicParameterParity}
\Gamma^{ \dot{5} } \, \epsilon_{r} ( x^{ \bar{\mu} }, x^{5} )
     = + \, \epsilon_{r} ( x^{\mu}, - x^{5} ).  
\end{equation}
}

\end{enumerate}
These conditions imply that the $ \mathbb{Z}_{2} $-odd quantities vanish 
at the orbifold fixed points.  
It is straightforward to check that $ S_{bulk} $ is 
$ \mathbb{Z}_{2} $-even and that the transformations
(\ref{transformations}) are consistent with the 
$ \mathbb{Z}_{2} $ parity assignments.

\section{\label{Calculation}Calculation of
$ \delta_{\epsilon} S_{bulk} $}

The $ D \! = \! 5 $ Chern-Simons AdS supergravity action is invariant (up 
to a boundary term) under its fermionic symmetry transformations.
In Section \ref{Bulk}, we modified this action and its fermionic 
transformations to obtain a $ \mathbb{Z}_{2} $-invariant bulk theory.
Due to the signum functions introduced by the modifications, 
$ \delta_{\epsilon} S_{bulk} $ contains terms which have no counterpart in 
the unmodified theory.
More specifically, the extra terms arise from $ \partial_{5} $ acting on 
the signum functions to yield delta functions.
Such `delta function' contributions to $ \delta_{\epsilon} S_{bulk} $
can potentially spoil the fermionic symmetry only at the 
$ \mathbb{Z}_{2} $ fixed points.
Thus, $ S_{bulk} $ is invariant under its 
fermionic transformations everywhere except perhaps at the 
$ \mathbb{Z}_{2} $ fixed points.
In this section, we will calculate $ \delta_{\epsilon} S_{bulk} $. 

For our metric 
and $ \mathbb{Z}_{2} $ parity assignments, 
the uncancelled variation $ \delta_{\epsilon} S_{bulk} $ arises from the 
variation of the 4-Fermi terms.  The 4-Fermi terms are
\begin{align}
\nonumber
S_{\psi^{4}}
  &= {\textstyle
       \frac{3i}{4} \int K 
       \left(
            \bar{\psi}^{r}_{\alpha} \psi^{\alpha}_{s}
            \bar{\psi}^{s}_{\beta} \nabla \psi^{\beta}_{r} 
          + \bar{\psi}^{s}_{\alpha} \bar{\psi}^{r}_{\beta}
            \psi^{\beta}_{s} \nabla \psi^{\alpha}_{r}   
       \right)
     + c.c.
     }
\\
\nonumber
  &= {\textstyle
       \frac{3i}{2} \int K 
       \bar{\psi}^{r}_{\alpha} \psi^{\alpha}_{s}
       \bar{\psi}^{s}_{\beta} \nabla \psi^{\beta}_{r}
     + c.c.
     }
\\
\nonumber
  &= {\textstyle
       \frac{3i}{2} \int K
       \left(
            \bar{\psi}^{r} \psi_{s}
       \right)  
       \left(
            \bar{\psi}^{s} \nabla \psi_{r}
       \right)
     + c.c.
     }          
\\
\nonumber
  &= {\textstyle
       \frac{3i}{2} \frac{1}{5!} \int d^{5}x \, 
       \varepsilon^{\mu \nu \rho \sigma \lambda} \,
       K 
       \left(
            \bar{\psi}^{r}_{\mu} \psi_{\nu s}
       \right)
       \left(
            \bar{\psi}^{s}_{\rho} \nabla_{\sigma} \psi_{\lambda r}
       \right)
     + c.c.
     }
\\
\nonumber
  &= {\textstyle
       \frac{3i}{2} \frac{1}{4!}
       \int d^{5}x \, K
       \left[
            \varepsilon^{ 5 \bar{\nu} \bar{\rho} \bar{\sigma} 
                          \bar{\lambda} }  
            \left(
                 \bar{\psi}^{r}_{5} \psi_{\bar{\nu} s}
            \right)
            \left(
                 \bar{\psi}^{s}_{\bar{\rho}} 
                 \nabla_{\bar{\sigma}} \psi_{\bar{\lambda} r}
           \right)
         + \varepsilon^{ \bar{\mu} 5 \bar{\rho} \bar{\sigma} 
                         \bar{\lambda} } 
            \left(
                 \bar{\psi}^{r}_{\bar{\mu}} \psi_{5s}
            \right)
            \left(
                 \bar{\psi}^{s}_{\bar{\rho}}
                 \nabla_{\bar{\sigma}} \psi_{\bar{\lambda} r}
           \right)    
       \right.
     }
\\
\nonumber
  &\quad \
     {\textstyle  
          + \varepsilon^{ \bar{\mu} \bar{\nu} 5 \bar{\sigma} 
                          \bar{\lambda} } 
            \left(
                 \bar{\psi}^{r}_{\bar{\mu}} \psi_{\bar{\nu} s}
            \right)
            \left(
                 \bar{\psi}^{s}_{5}
                 \nabla_{\bar{\sigma}} \psi_{\bar{\lambda} r}
           \right)
         + 
            \varepsilon^{ \bar{\mu} \bar{\nu} \bar{\rho} 5 \bar{\lambda} } 
            \left(
                 \bar{\psi}^{r}_{ \bar{\mu} } \psi_{ \bar{\nu} s }
            \right)
            \left(
                 \bar{\psi}^{s}_{ \bar{\rho} }
                 \nabla_{5} \psi_{ \bar{\lambda} r }
           \right)
     }
\\
\label{S4Fermi}
  &\quad \
     {\textstyle
          + \varepsilon^{ \bar{\mu} \bar{\nu} \bar{\rho} \bar{\sigma} 5 } 
            \left( 
                 \bar{\psi}^{r}_{ \bar{\mu} } \psi_{ \bar{\nu} s }
            \right)
            \left(
                 \bar{\psi}^{s}_{ \bar{\rho} }
                 \nabla_{ \bar{\sigma} } \psi_{5r}
           \right)
       \Bigl.
       \Bigr]
     + c.c.     
     }  
\end{align}
Let us now compute $ \delta_{\epsilon} S_{bulk} $ by applying
$ \delta_{\epsilon} $ to (\ref{S4Fermi}) and dropping all terms which 
contribute no delta functions.
For our metric and $ \mathbb{Z}_{2} $ parity assignments, we can drop 
all but
\begin{enumerate}

\item{The $ \partial_{\mu} $ part of $ \nabla_{\mu} $. }

\item{The $ - \partial_{\mu} \epsilon_{r} $ part of 
$ \delta_{\epsilon} = - \nabla_{\mu} \epsilon_{r} $. }     

\end{enumerate}
The only contributing terms are thus contained in the expression
\begin{align}
\nonumber
Q &\equiv
     {\textstyle 
     - \frac{3i}{2} \frac{1}{4!}
       \int d^{5}x \, K
     \left\{
       \varepsilon^{ 5 \bar{\nu} \bar{\rho} \bar{\sigma} \bar{\lambda} }        
       \left(    
                 \partial_{5} \bar{\epsilon}^{r}
            \psi_{\bar{\nu} s}
       \right)
       \left(
            \bar{\psi}^{s}_{ \bar{\rho} } 
            \partial_{ \bar{\sigma} } \psi_{ \bar{\lambda} r } 
       \right)
     \right.
     }
\\
\nonumber
  &\quad \
     {\textstyle
     + \varepsilon^{ \bar{\mu} 5 \bar{\rho} \bar{\sigma} \bar{\lambda} } 
       \left(
            \bar{\psi}^{r}_{ \bar{\mu} }
            \partial_{5} \epsilon_{s}
       \right)
       \left(
            \bar{\psi}^{s}_{ \bar{\rho} }
            \partial_{ \bar{\sigma} } \psi_{ \bar{\lambda} r }
       \right)
     + \varepsilon^{ \bar{\mu} \bar{\nu} 5 \bar{\sigma} \bar{\lambda} }
       \left(
            \bar{\psi}^{r}_{ \bar{\mu} }
            \psi_{ \bar{\nu} s }
       \right)
       \left( 
            \partial_{5} \bar{\epsilon}^{s}
            \partial_{ \bar{\sigma} } \psi_{ \bar{\lambda} r }
       \right)
     }
\\
\nonumber
  &\quad \
     {\textstyle
     + \varepsilon^{ \bar{\mu} \bar{\nu} \bar{\rho} 5 \bar{\lambda} }
       \Bigl[
            \left(
                 \partial_{ \bar{\mu} } \bar{\epsilon}^{r}
                 \psi_{ \bar{\nu} s }
            \right)
            \left(
                 \bar{\psi}^{s}_{ \bar{\rho} } 
                 \partial_{5} \psi_{ \bar{\lambda} r } 
            \right)
          + \left(
                 \bar{\psi}^{r}_{ \bar{\mu} } 
                 \partial_{ \bar{\nu} } \epsilon_{s}
            \right)
            \left(
                 \bar{\psi}^{s}_{ \bar{\rho} }
                 \partial_{5} \psi_{ \bar{\lambda} r }
            \right)
       \Bigr.
     }
\\
\nonumber 
  &\quad \
     {\textstyle
          \phantom{ \varepsilon^{ \bar{\mu} \bar{\nu} \bar{\rho} 5 
                    \bar{\lambda} } }
          + \left(
                 \bar{\psi}^{r}_{ \bar{\mu} } 
                 \psi_{ \bar{\nu} s }
            \right)
            \left( 
                 \partial_{ \bar{\rho} } \bar{\epsilon}^{s}
                 \partial_{5} \psi_{ \bar{\lambda} r }
            \right)
          + \left(
                 \bar{\psi}^{r}_{ \bar{\mu} }
                 \psi_{ \bar{\nu} s }               
            \right)
            \left(
                 \bar{\psi}^{s}_{ \bar{\rho} }
                 \partial_{5} \partial_{ \bar{\lambda} } \epsilon_{r}
            \right)
        \Bigl.
        \Bigr]
     }
\\
  &\quad \
    {\textstyle
     + \varepsilon^{ \bar{\mu} \bar{\nu} \bar{\rho} \bar{\sigma} 5 } \, 
            \left(
                 \bar{\psi}^{r}_{ \bar{\mu} }
                 \psi_{ \bar{\nu} s }
            \right)
            \left(
                 \bar{\psi}^{s}_{ \bar{\rho} }
                 \partial_{ \bar{\sigma} } \partial_{5 } \epsilon_{r}
            \right)
        \Bigl.
        \Bigr\}
     + c.c.
     }  
\end{align}
More specifically, the delta function terms contained in $ Q $ are 
obtained by integrating by parts and keeping only the terms in which 
$ \partial_{5} $ acts on $ K $.
Thus,
\begin{align}
\nonumber
\delta_{\epsilon} S_{bulk} &=
     {\textstyle
     \frac{3i}{2} \frac{1}{4!}
       \int d^{5}x \, \partial_{5} K
     \left\{
       \varepsilon^{ 5 \bar{\nu} \bar{\rho} \bar{\sigma} \bar{\lambda} }
       \left(
            \bar{\epsilon}^{r}
            \psi_{\bar{\nu} s}
       \right)
       \left(
            \bar{\psi}^{s}_{ \bar{\rho} }
            \partial_{ \bar{\sigma} } \psi_{ \bar{\lambda} r }
       \right)  
     \right.
     }
\\
\nonumber
  &\quad \
     {\textstyle   
     + \varepsilon^{ \bar{\mu} 5 \bar{\rho} \bar{\sigma} \bar{\lambda} } 
       \left(
            \bar{\psi}^{r}_{ \bar{\mu} }
            \epsilon_{s}
       \right)
       \left(
            \bar{\psi}^{s}_{ \bar{\rho} }
            \partial_{ \bar{\sigma} } \psi_{ \bar{\lambda} r }
       \right)
     + \varepsilon^{ \bar{\mu} \bar{\nu} 5 \bar{\sigma} \bar{\lambda} }
       \left(
            \bar{\psi}^{r}_{ \bar{\mu} }
            \psi_{ \bar{\nu} s }
       \right)
       \left(
            \bar{\epsilon}^{s}
            \partial_{ \bar{\sigma} } \psi_{ \bar{\lambda} r }
       \right)     
     }
\\
\nonumber
  &\quad \
     {\textstyle
     + \varepsilon^{ \bar{\mu} \bar{\nu} \bar{\rho} 5 \bar{\lambda} }
       \Bigl[
            \left(
                 \partial_{ \bar{\mu} } \bar{\epsilon}^{r}
                 \psi_{ \bar{\nu} s }
            \right)
            \left(
                 \bar{\psi}^{s}_{ \bar{\rho} }
                 \psi_{ \bar{\lambda} r }
            \right)
          + \left(
                 \bar{\psi}^{r}_{ \bar{\mu} }
                 \partial_{ \bar{\nu} } \epsilon_{s}
            \right)
            \left(
                 \bar{\psi}^{s}_{ \bar{\rho} }
                 \psi_{ \bar{\lambda} r }
            \right)
       \Bigr.
     }
\\
\nonumber
  &\quad \
     {\textstyle
          \phantom{ \varepsilon^{ \bar{\mu} \bar{\nu} \bar{\rho} 5
                    \bar{\lambda} } }
          + \left(
                 \bar{\psi}^{r}_{ \bar{\mu} }
                 \psi_{ \bar{\nu} s }
            \right)
            \left(
                 \partial_{ \bar{\rho} } \bar{\epsilon}^{s}
                 \psi_{ \bar{\lambda} r }
            \right)
          + \left(
                 \bar{\psi}^{r}_{ \bar{\mu} } 
                 \psi_{ \bar{\nu} s }
            \right)
            \left(
                 \bar{\psi}^{s}_{ \bar{\rho} }
                 \partial_{ \bar{\lambda} } \epsilon_{r}
            \right)
        \Bigl.
        \Bigr]
     }
\\
\label{delta_epsilonS_bulk}
 &\quad \
    {\textstyle
     + \varepsilon^{ \bar{\mu} \bar{\nu} \bar{\rho} \bar{\sigma} 5 } 
            \left(
                 \bar{\psi}^{r}_{ \bar{\mu} }
                 \psi_{ \bar{\nu} s }
            \right)
            \left(
                 \bar{\psi}^{s}_{ \bar{\rho} }
                 \partial_{ \bar{\sigma} } \epsilon_{r}
            \right)
        \Bigl.     
        \Bigr\}
     + c.c.,  
     }
\end{align}
where
\begin{equation}
\label{partial_5K}
\partial_{5} K 
  = 2k \left[ 
            \delta( x^{5} ) - \delta( x^{5} - \pi \rho ) 
       \right]. 
\end{equation}

\section{\label{ModifiedBulk}Modified $ \mathbb{Z}_{2} $-invariant bulk 
theory}

The result (\ref{delta_epsilonS_bulk}) for $ \delta_{\epsilon} S_{bulk} $ 
demonstrates that $ S_{bulk} $  is not invariant under the fermionic 
transformations $ \delta_{\epsilon} $.
In this section, we will modify 
$ S_{bulk} $ and $ \delta_{\epsilon} $
by replacing the 
$ adS_{5} \times su(N) \times u(1) $ covariant derivative 
$ \nabla $ with $ \widetilde{\nabla} $, where
\begin{align}    
\nonumber
\widetilde{\nabla}_{\sigma} \psi_{\lambda r}
  &\equiv   \nabla_{\sigma} \psi_{\lambda r}
          + 2 \delta^{5}_{\sigma} \delta^{ \bar{\lambda} }_{\lambda}   
            \left[
                 \delta( x^{5} ) - \delta( x^{5} - \pi \rho )
            \right]
            \mathrm{sgn}(x^{5}) \Gamma_{ \dot{5} }
            \psi_{ \bar{\lambda} r },
\\
\label{ModifiedNabla}
\widetilde{\nabla}_{\sigma} \epsilon_{r}
  &\equiv   \nabla_{\sigma} \epsilon_{r}
          + 2 \delta^{5}_{\sigma}
            \left[
                 \delta( x^{5} ) - \delta( x^{5} - \pi \rho )
            \right]
            \mathrm{sgn}(x^{5}) \Gamma_{ \dot{5} } \epsilon_{r}.
\end{align}
We will show that the modified bulk action
\begin{equation}
\tilde{S}_{bulk} 
   \equiv S_{bulk}
          \bigl(
               \nabla \rightarrow \widetilde{\nabla}
          \bigr)
   \equiv S_{bulk} + \Delta S_{bulk}
\end{equation}
is invariant under the modified transformations
\begin{equation}
\tilde{\delta}_{\epsilon} 
   \equiv \delta_{\epsilon}
          \bigl(
               \nabla \rightarrow \widetilde{\nabla}
          \bigr)
   \equiv \delta_{\epsilon} + \Delta \delta_{\epsilon}.
\end{equation}
That is, we will show that
\begin{equation}
\label{TildeDelta_epsilonS_bulkDefinition}
\tilde{\delta}_{\epsilon} \tilde{S}_{bulk}
   =   \delta_{\epsilon} S_{bulk} 
     + \left( 
            \Delta \delta_{\epsilon} 
       \right) S_{bulk}
     + \tilde{\delta}_{\epsilon} 
       \left( 
            \Delta S_{bulk}
       \right)  
\end{equation}
vanishes.
It is straightforward to check that $ \tilde{S}_{bulk} $ is 
$ \mathbb{Z}_{2} $-invariant and the transformations 
$ \tilde{\delta}_{\epsilon} $ are consistent with our 
$ \mathbb{Z}_{2} $ parity assignments.

We begin by computing
$ \left( \Delta \delta_{\epsilon} \right) S_{bulk} $.
For our metric and $ \mathbb{Z}_{2} $ parity assignments, the only part of
$ S_{bulk} $ which is not invariant under
$ \Delta \delta_{\epsilon} $  is 
$ S_{\psi^{4}} $ (given by (\ref{S4Fermi})).
Note that
\begin{equation}
\left( 
     \Delta \delta_{\epsilon} 
\right) 
\psi_{\lambda r}
   = -  2 \delta^{5}_{\lambda}
        \left[
             \delta( x^{5} ) - \delta( x^{5} - \pi \rho )
        \right]
        \mathrm{sgn}(x^{5}) \Gamma_{ \dot{5} } \epsilon_{r}.
\end{equation}
Thus, after using $ K \, \mathrm{sgn}( x^{5} ) = k $,
(\ref{FermionicParameterParity}), and (\ref{partial_5K}), we obtain
\begin{align}
\nonumber
\left( \Delta \delta_{\epsilon} \right) S_{bulk}
  &= {\textstyle
     - \frac{3i}{2} \frac{1}{4!} \int d^{5}x \, \partial_{5} K \,
       \left[
            \varepsilon^{ 5 \bar{\nu} \bar{\rho} \bar{\sigma} 
                          \bar{\lambda} }
            \left(
                 \bar{\epsilon}^{r} 
                 \psi_{ \bar{\nu} s }         
            \right)
            \left( 
                 \bar{\psi}^{s}_{ \bar{\rho} } 
                 \partial_{ \bar{\sigma} } \psi_{ \bar{\lambda} r }
            \right) 
       \right.
     }
\\
\nonumber
  &\quad \   
     {\textstyle  
     + \varepsilon^{ \bar{\mu} 5 \bar{\rho} \bar{\sigma}
                       \bar{\lambda} }
            \left(
                 \bar{\psi}^{r}_{ \bar{\mu} }  
                 \epsilon_{s}
            \right)
            \left(
                 \bar{\psi}^{s}_{ \bar{\rho} }
                 \partial_{ \bar{\sigma} } \psi_{ \bar{\lambda} r }
            \right)
     + \varepsilon^{ \bar{\mu} \bar{\nu} 5  \bar{\sigma} \bar{\lambda} }
            \left(
                 \bar{\psi}^{r}_{ \bar{\mu} }
                 \psi_{ \bar{\nu} s } 
            \right)
            \left(
                 \bar{\epsilon}^{s}
                 \partial_{ \bar{\sigma} } \psi_{ \bar{\lambda} r }
            \right)  
     }
\\
\label{Deltadelta_epsilonS_bulk}
&\quad \
     {\textstyle
     + \varepsilon^{ \bar{\mu} \bar{\nu} \bar{\rho} \bar{\sigma} 5 }
            \left(
                 \bar{\psi}^{r}_{ \bar{\mu} }
                 \psi_{ \bar{\nu} s }
            \right)
            \left(
                 \bar{\psi}^{s}_{ \bar{\rho} }
                 \partial_{ \bar{\sigma} } \epsilon_{r}
            \right)
     \Bigl.
     \Bigr]
     + c.c.
     }
\end{align}

Now, let us compute
$ \tilde{\delta}_{\epsilon} \left( \Delta S_{bulk} \right) $.
For our metric and $ \mathbb{Z}_{2} $ parity assignments, the only part of
$ S_{bulk} $ which is changed by the replacement
$ \nabla \rightarrow \widetilde{\nabla} $ is
$ S_{\psi^{4}} $.
After using $ K \, \mathrm{sgn}( x^{5} ) = k $,
(\ref{GravitinoParity}), and (\ref{partial_5K}), we obtain
\begin{equation}  
\label{DeltaS_bulk}
\Delta S_{bulk}
   = {\textstyle   
       \frac{3i}{2} \frac{1}{4!} \int d^{5}x \, \partial_{5} K \,
       \varepsilon^{ \bar{\mu} \bar{\nu} \bar{\rho} 5 \bar{\lambda} }
       \left(
            \bar{\psi}^{r}_{ \bar{\mu} } \psi_{ \bar{\nu} s }
       \right)
       \left(
            \bar{\psi}^{s}_{ \bar{\rho} } \psi_{ \bar{\lambda} r }
       \right)
     + c.c.
     }
\end{equation}
Applying $ \tilde{\delta}_{\epsilon} $ to (\ref{DeltaS_bulk}) yields
\begin{align}
\nonumber
\tilde{\delta}_{\epsilon} \left( \Delta S_{bulk} \right)
  &= {\textstyle
     - \frac{3i}{2} \frac{1}{4!} \int d^{5}x \, \partial_{5} K \,
       \varepsilon^{ \bar{\mu} \bar{\nu} \bar{\rho} 5 \bar{\lambda} }
       \Bigl[
            \left(
                 \partial_{ \bar{\mu} } \bar{\epsilon}^{r}
                 \psi_{ \bar{\nu} s }
            \right)
            \left(
                 \bar{\psi}^{s}_{ \bar{\rho} }
                 \psi_{ \bar{\lambda} r }
            \right)
          + \left(
                 \bar{\psi}^{r}_{ \bar{\mu} }
                 \partial_{ \bar{\nu} } \epsilon_{s}
            \right)
            \left(
                 \bar{\psi}^{s}_{ \bar{\rho} }
                 \psi_{ \bar{\lambda} r }
            \right)
       \Bigr.
     }
\\
\label{tildedelta_epsilonDeltaS_bulk}
  &\quad \
      {\textstyle
          + \left(
                 \bar{\psi}^{r}_{ \bar{\mu} }
                 \psi_{ \bar{\nu} s }
            \right)
            \left(
                 \partial_{ \bar{\rho} } \bar{\epsilon}^{s}
                 \psi_{ \bar{\lambda} r }
            \right)
          + \left( 
                 \bar{\psi}^{r}_{ \bar{\mu} } 
                 \psi_{ \bar{\nu} s }
            \right)
            \left(
                 \bar{\psi}^{s}_{ \bar{\rho} }
                 \partial_{ \bar{\lambda} } \epsilon_{r}
            \right)
        \Bigl.
        \Bigr]
      + c.c.
      }
\end{align}

Using the results 
(\ref{delta_epsilonS_bulk}),
(\ref{Deltadelta_epsilonS_bulk}), and 
(\ref{tildedelta_epsilonDeltaS_bulk}) in
(\ref{TildeDelta_epsilonS_bulkDefinition}) yields
\begin{equation}
\tilde{\delta}_{\epsilon} \tilde{S}_{bulk} = 0.
\end{equation}

\section{\label{Brane}Brane action}

To complete our model, we add the brane action $ S_{brane} $.
In the absence of particle excitations, the brane action consists of brane 
tensions.
That is,
\begin{equation}
\label{S_brane}
S_{brane}
   = - \int d^{5}x \, e^{(4)}
       \left[
            \mathcal{T}^{ (0) } \delta( x^{5} )
          + \mathcal{T}^{ ( \pi \rho ) } \delta( x^{5} - \pi \rho )
       \right] + \mathrm{excitations},
\end{equation}
where 
$ e^{(4)} \equiv \mathrm{det} ( e_{ \bar{\mu} }{}^{ \bar{a} } ) $.
As discussed in Section \ref{Bulk}, $ \mathbb{Z}_{2} $-odd quantities 
vanish at the $ \mathbb{Z}_{2} $ fixed points.
Thus, it is clear that $ S_{brane} $ is $ \mathbb{Z}_{2} $-even.
Further discussion of 3-brane actions can be found in 
\cite{Sundrum:Effective}.

\section{\label{Connection}Connection with original RS model}

In this section, we will show that our total action
$ S = \tilde{S}_{bulk} + S_{brane} $
reduces to the original Randall-Sundrum model \cite{RandSund:ALarge}
when $ \tilde{S}_{bulk} $ is restricted to its gravitational sector.

The gravitational sector of $ \tilde{S}_{bulk} $ is $ S_{grav} $, given by 
the first equation of (\ref{S_bulkSectors}).
$ S_{grav} $ consists of three terms: 
\begin{enumerate}

\item{The `Gauss-Bonnet' term
      $ \int \frac{1}{8} K \varepsilon_{abcde}
         R^{ab} R^{cd} e^{e} / L $. }

\item{The `Einstein-Hilbert' term
      $ \int \frac{1}{8} \cdot \frac{2}{3} 
        K \varepsilon_{abcde}
        R^{ab} e^{c} e^{d} e^{e} / L^{3} $. }

\item{The `cosmological constant' term
      $ \int \frac{1}{8} \cdot \frac{1}{5} 
        K \varepsilon_{abcde}
        e^{a} e^{b} e^{c} e^{d} e^{e} / L^{5} $. }

\end{enumerate}
For our metric, the first term can be expressed as a linear combination of 
the other two.
Summing the three terms yields an effective Einstein-Hilbert term and 
an effective cosmological constant term.
To demonstrate this explicitly, let us evaluate $ S_{grav} $ for our 
metric.
Using the results in the Appendix, we obtain
\begin{align}
\nonumber
\varepsilon_{abcde} R^{ab} R^{cd} e^{e}
  &= d^{5}x \; e
     \left(
          - \frac{120}{ \ell^{4} } 
        + \frac{192}{ \ell^{3} } 
          \left[
               \delta( x^{5} ) - \delta( x^{5} - \pi \rho )
          \right]
     \right),
\\
\nonumber
\varepsilon_{abcde} R^{ab} e^{c} e^{d} e^{e}
  &= d^{5}x \; e \left( -6R \right),
\\
\varepsilon_{abcde} e^{a} e^{b} e^{c} e^{d} e^{e}
  &= d^{5}x \; e \left( -5! \right),
\end{align}
where 
$ e \equiv \mathrm{det} ( e_{\mu}{}^{a} ) $.
Thus,
\begin{align}
\nonumber
S_{grav} 
  &= \int d^{5}x \; e \, \frac{1}{8}
     \left\{
          \frac{k}{ \ell }
          \left(   
             - \frac{120}{ \ell^{4} }
             + \frac{192}{ \ell^{3} }
               \left[
                    \delta( x^{5} ) - \delta( x^{5} - \pi \rho )
               \right]
          \right)
     \right. 
\\
\nonumber
  &\quad \,
     \left.     
        + \frac{2k}{ 3 \ell^{3} }
          \left(
             - 6R
          \right)
        + \frac{k}{ 5 \ell^{5} } 
          \left( 
             - 5! 
          \right)
     \right\}   
\\
\nonumber
  &= \int d^{5}x \; e \, \frac{k}{ \ell^{3} }
     \left\{
        - \frac{15}{ \ell^{2} }
        + \frac{24}{\ell}
          \left[
               \delta( x^{5} ) - \delta( x^{5} - \pi \rho )
          \right]
        - \frac{1}{2} R
        - \frac{3}{ \ell^{2} }
     \right\}
\\
\nonumber
  &= \int d^{5}x \; e \, \frac{k}{ \ell^{3} }
     \left\{
          \frac{3}{2}
          \left(
             - \frac{20}{ \ell^{2} } 
             + \frac{16}{\ell}
               \left[
                    \delta( x^{5} ) - \delta( x^{5} - \pi \rho )
               \right] 
          \right)
        - \frac{1}{2} R
        + \frac{12}{ \ell^{2} }
     \right\}
\\
\nonumber
  &= \int d^{5}x \; e \, \frac{k}{ \ell^{3} }
     \left(
          R + \frac{12}{ \ell^{2} }
     \right)
\\
\label{S_gravOurMetric}
  &= \int d^{5}x \; e 
     \left(
          2 M^{3} R - \Lambda
     \right), 
\end{align}
where $ M $ is the five-dimensional gravitational mass scale\footnote
{
$ M $ is related to the four-dimensional gravitational mass scale
$ M_{(4)} = 2.43 \times 10^{18} $ GeV by
$ M^{2}_{(4)}
   = M^{3} \int_{- \pi \rho}^{+ \pi \rho} dx^{5} \, 
     \mathfrak{a}^{2}(x^{5})
   = M^{3} \ell \left[
                     1 - \exp(-2 \pi \rho / \ell)
                \right] $.
The effective mass scales on the 3-branes at
$ x^{5} = 0 $ and $ x^{5} = \pi \rho $ are respectively
$ M_{(4)} $ and $ M_{(4)} e^{ - \pi \rho / \ell } $.
If the Standard Model fields live on the 3-brane at
$ x^{5} = \pi \rho $, then $ M_{(4)} e^{ - \pi \rho / \ell} $ can be
associated with the electroweak scale.
},
$ \Lambda $ is the bulk cosmological constant, and
\begin{equation}
M^{3} = \frac{k}{ 2 \ell^{3} },
\quad
\Lambda = - \frac{ 24 M^{3} }{ \ell^{2} }. 
\end{equation}

Combining the result (\ref{S_gravOurMetric}) with (\ref{S_brane}) , we 
obtain the action of the original Randall-Sundrum model.
It is shown in \cite{RandSund:ALarge} that the five-dimensional vacuum 
Einstein's equations for this system,
\begin{equation}
R_{\mu \nu} - \frac{1}{2} g_{\mu \nu} R
 = - \frac{1}{4 M^{3} }
     \left\{
          g_{\mu \nu} \Lambda 
        + \frac{ e^{(4)} }{e} 
          \delta^{ \bar{\mu} }_{\mu}
          \delta^{ \bar{\nu} }_{ \nu} 
          g_{ \bar{\mu} \bar{\nu} }
          \left[
               \mathcal{T}^{(0)} \delta ( x^{5} )
             + \mathcal{T}^{(\pi \rho)} \delta ( x^{5} - \pi \rho )
          \right]
     \right\},
\end{equation}
are solved by our metric provided that the relations
\begin{equation}
\label{fine-tuning}
\mathcal{T}^{(0)} = - \mathcal{T}^{(\pi \rho)}
                  =   24 M^{3} / \ell ,
\quad
\Lambda = {\textstyle
          - 24 M^{3} / \ell^{2}
          }
\end{equation}
are satisfied.

\section{\label{Killing}Killing spinors}

In this section, we will solve the Killing spinor equations for a purely 
bosonic configuration with vanishing $ su(N) $ and $ u(1) $ gauge fields.
In this case, the Killing spinor equations reduce to
\begin{align}
\nonumber
0 &=   \delta_{\epsilon} \psi_{ \bar{\mu} r } 
   = - \partial_{ \bar{\mu} } \epsilon_{r}
     - \frac{1}{2} 
       \frac{ \mathfrak{a}^{\prime} }{ \mathfrak{a} } 
       \Gamma_{ \bar{\mu} } 
       \left( 
            \Gamma_{ \dot{5} } - 1
       \right)
       \epsilon_{r},
\\
0 &= \delta_{\epsilon} \psi_{5 r}
   = - \partial_{5} \epsilon_{r}
     + \frac{1}{2}         
       \frac{ \mathfrak{a}^{\prime} }{ \mathfrak{a} } 
       \Gamma_{ \dot{5} } \epsilon_{r}
     - 2 \left[
              \delta( x^{5} ) - \delta( x^{5} - \pi \rho)
         \right]
         \mathrm{sgn}( x^{5} ) \Gamma_{ \dot{5} } \epsilon_{r}.
\end{align}       
To solve these equations, split $ \epsilon_{r} $ into 
$ \mathbb{Z}_{2} $-even  $ ( \epsilon^{+}_{r} ) $ and 
$ \mathbb{Z}_{2} $-odd  $ ( \epsilon^{-}_{r} ) $ pieces:
\begin{equation}
\epsilon_{r} = \epsilon^{+}_{r} + \epsilon^{-}_{r},
\end{equation}
where
\begin{equation}
\epsilon^{\pm}_{r} 
   \equiv {\textstyle
          \frac{1}{2}
          }
          \left(
               \epsilon_{r} \pm \Gamma_{ \dot{5} } \epsilon_{r}
          \right)
   = \pm \Gamma_{ \dot{5} } \epsilon^{\pm}_{r}.
\end{equation}
We obtain the following system of equations:
\begin{align}
\nonumber
\partial_{ \bar{\mu} } \epsilon^{+}_{r}
  &= - \left( 
            \mathfrak{a}^{\prime} / \mathfrak{a} 
       \right)
       \Gamma_{ \bar{\mu} } \Gamma_{ \dot{5} } \epsilon^{-}_{r},
\\
\nonumber
\partial_{ \bar{\mu} } \epsilon^{-}_{r}
  &= 0,
\\
\nonumber
\partial_{5} \epsilon^{+}_{r}
  &= + {\textstyle
       \frac{1}{2}
       } 
       \left( 
            \mathfrak{a}^{\prime} / \mathfrak{a}
       \right)
       \epsilon^{+}_{r} 
     - 2 
       \left[
            \delta( x^{5} ) - \delta( x^{5} - \pi \rho)
       \right]
       \mathrm{sgn}( x^{5} ) \epsilon^{+}_{r},
\\
\label{system}
\partial_{5} \epsilon^{-}_{r}
  &= - {\textstyle 
       \frac{1}{2}
       } 
       \left(
            \mathfrak{a}^{\prime} / \mathfrak{a}
       \right)
       \epsilon^{-}_{r}
     + 2 
       \left[
            \delta( x^{5} ) - \delta( x^{5} - \pi \rho)
       \right]
       \mathrm{sgn}( x^{5} ) \epsilon^{-}_{r}.
\end{align}         
These equations are solved by
\begin{align}
\nonumber
\epsilon^{+}_{r} 
  &= \mathfrak{a}^{1/2}
     \left[
          - \left( 
                 \mathfrak{a}^{\prime} / \mathfrak{a}^{2} 
            \right) 
            x^{ \bar{\mu} } \Gamma_{ \bar{\mu} }
            \Gamma_{ \dot{5} } \, \mathrm{sgn}( x^{5} ) \,
            \chi^{-(0)}_{r} 
          + \chi^{+(0)}_{r}
     \right]
\\
\nonumber
  &= \mathfrak{a}^{1/2}
     \left[ 
            \left( 1 / \ell \right) 
            x^{ \bar{\mu} } 
            \delta_{ \bar{\mu} }{}^{ \bar{a} } \Gamma_{ \bar{a} }
            \Gamma_{ \dot{5} } \, \chi^{-(0)}_{r}
          + \chi^{+(0)}_{r}
     \right],
\\
\label{EvenOddSolutions}
\epsilon^{-}_{r}
  &= \mathfrak{a}^{-1/2} \, \mathrm{sgn}( x^{5} ) \, \chi^{-(0)}_{r}, 
\end{align}
where $ \chi^{+(0)}_{r} $ and $ \chi^{-(0)}_{r} $ are constant (projected)
spinors.\footnote
{
It is straightforward to check that (\ref{EvenOddSolutions}) satisfies
the first, second, and fourth equations of (\ref{system}).
There is, however, a subtlety in checking that (\ref{EvenOddSolutions}) 
satisfies the third equation of (\ref{system}).
Unlike $ \epsilon^{-}_{r} $, $ \epsilon^{+}_{r} $ is a smooth function of 
$ x^{5} $. Thus, the second term on the right side of the third 
equation of (\ref{system}) contributes nothing.
}
Thus, our solution for the Killing spinors is
\begin{equation}
\epsilon_{r}
   =   \mathfrak{a}^{1/2} \chi^{+(0)}_{r}
     + \mathfrak{a}^{-1/2} \, \mathrm{sgn}( x^{5} )
       \left(
            1 
          - \frac{ \mathfrak{a}^{\prime} }{ \mathfrak{a} }
            x^{ \bar{\mu} } \Gamma_{ \bar{\mu} }
            \Gamma_{ \dot{5} }
       \right)
       \chi^{-(0)}_{r}.
\end{equation}

\section{\label{Conclusion}Conclusion}

We  have constructed a Randall-Sundrum scenario from 
$ D \! = \! 5 $ Chern-Simons AdS supergravity.
Our total action $ S = \tilde{S}_{bulk} + S_{brane} $ is 
$ \mathbb{Z}_{2} $-invariant.
$ \tilde{S}_{bulk} $ is invariant under the fermionic transformations 
$ \tilde{\delta}_{\epsilon} $.
However,
\begin{equation}
\tilde{\delta}_{\epsilon} S_{brane}
= - \int d^{5}x \, 
         \tilde{\delta}_{\epsilon}e^{(4)}
       \left[
            \mathcal{T}^{ (0) } \delta( x^{5} )
          + \mathcal{T}^{ ( \pi \rho ) } \delta( x^{5} - \pi \rho )
       \right] + \cdots,   
\end{equation}
where
\begin{equation}
\tilde{\delta}_{\epsilon}e^{(4)}
   = {\textstyle
     e^{(4)}
     \left[
        - \frac{1}{2}
          \left(
               \bar{\psi}^{r}_{ \bar{\mu} }
               \Gamma^{ \bar{\mu} }
               \epsilon_{r}
             - \bar{\epsilon}^{r}
               \Gamma^{ \bar{\mu} }
               \psi_{ \bar{\mu} r }
          \right)
     \right].
     }
\end{equation} 
Thus, the fermionic symmetry is broken by $ S_{brane} $.
Nevertheless, the Killing spinors of Section \ref{Killing} are globally 
defined. 

Our model reduces to the original Randall-Sundrum model 
\cite{RandSund:ALarge}
when $ \tilde{S}_{bulk} $ is restricted to its gravitational sector.
The original Randall-Sundrum model involves the fine-tuning relations
\begin{equation*}  
\mathcal{T}^{(0)} = - \mathcal{T}^{(\pi \rho)}
                  =   24 M^{3} / \ell ,
\quad
\Lambda = {\textstyle
          - 24 M^{3} / \ell^{2}.
          }
\end{equation*}
Randall-Sundrum scenarios constructed from standard
$ D \! = \! 5 $ supergravity theories yield these relations 
(up to an overall normalization factor) as a consequence of local 
supersymmetry (some examples are given in \cite{SUSYRS}). 
In our case, the relation 
$ \Lambda = - 24 M^{3} / \ell^{2} $
follows from our metric choice.
We do not obtain the relations 
$ \mathcal{T}^{(0)} = - \mathcal{T}^{(\pi \rho)}
                    =   24 M^{3} / \ell $
as a consequence of local fermionic symmetry.
These are fine-tuning relations in our model.

\appendix
\section{\label{Appendix}Appendix}

In this Appendix, we work out the f\"{u}nfbein, spin connection,
curvature 2-form components, Ricci tensor, and Ricci scalar for our
metric (\ref{metric}).
For the f\"{u}nfbein, we obtain
\begin{align}
\nonumber
e_{ \bar{\mu} }{}^{ \bar{a} }
  &= \mathfrak{a} \delta_{ \bar{\mu} }{}^{ \bar{a} },
\quad
e_{ \bar{\mu} \bar{a} }
   = e_{ \bar{\mu} }{}^{ \bar{b} } \eta_{ \bar{b} \bar{a} },
\quad
e^{ \bar{\mu} \bar{a} }
   = g^{ \bar{\mu} \bar{\nu} } e_{ \bar{\nu} }{}^{ \bar{a} },
\\
\nonumber
e_{ \bar{a} }{}^{ \bar{\mu} }
  &= \mathfrak{a}^{-1} \delta_{ \bar{a} }{}^{ \bar{\mu} },
\quad
e_{ \bar{a} \bar{\mu} }
   = e_{ \bar{a} }{}^{ \bar{\nu} } g_{ \bar{\nu} \bar{\mu} },
\quad
e^{ \bar{a} \bar{\mu} }
   = \eta^{ \bar{a} \bar{b} } e_{ \bar{b} }{}^{ \bar{\mu} },
\\
e_{5}{}^{ \dot{5} }
  &= e_{ 5 \dot{5} } =  e^{ 5 \dot{5} } = 1,
\quad
e_{ \dot{5} }{}^{5}
   = e_{ \dot{5} 5 } = e^{ \dot{5} 5 } = 1.
\end{align}
Our conventions for the spin connection, curvature 2-form
components, Ricci tensor, and Ricci scalar are respectively
\begin{align*}
\omega_{\mu}{}^{ab}
  &= \textstyle{ \frac{1}{2} } e^{\nu a}
          ( \partial_{\mu} e_{\nu}{}^{b} - \partial_{\nu} e_{\mu}{}^{b} )
     - \textstyle{ \frac{1}{2} } e^{\nu b}
          ( \partial_{\mu} e_{\nu}{}^{a} - \partial_{\nu} e_{\mu}{}^{a} )
\\
  &\quad
     - \textstyle{ \frac{1}{2} } e^{\rho a} e^{\sigma b}
          ( \partial_{\rho} e_{\sigma c} - \partial_{\sigma} e_{\rho c} )
          e^{c}{}_{\mu},
\\
R_{\mu \nu}{}^{ab}
  &=   \partial_{\mu} \omega_{\nu}{}^{ab}
     - \partial_{\nu} \omega_{\mu}{}^{ab}
     + \omega_{\mu}{}^{ac} \omega_{\nu c}{}^{b}  
     - \omega_{\nu}{}^{ac} \omega_{\mu c}{}^{b},
\\
R_{\nu \sigma}
  &= R_{\mu \nu}{}^{ab} e_{a}{}^{\mu} e_{b \sigma},
\quad
R  = e_{a}{}^{\mu} e_{b}{}^{\nu} R_{\mu \nu}{}^{ab}.
\end{align*}
For the metric (\ref{metric}), the nonzero quantities here are
\begin{equation}
\omega_{ \bar{\mu} }{}^{ \bar{a} \dot{5} }
  = - \omega_{ \bar{\mu} }{}^{ \dot{5} \bar{a} }
  = \mathfrak{a}^{\prime} \delta_{ \bar{\mu} }{}^{ \bar{a} }
  = - e_{ \bar{\mu} }{}^{ \bar{a} } / L,
\end{equation}
\begin{gather}
\nonumber
R_{ \bar{\mu} \bar{\nu} }{}^{ \bar{a} \bar{b} }
   = - \mathfrak{a}^{\prime \, 2}
       (
          \delta_{ \bar{\mu} }{}^{ \bar{a} }
          \delta_{ \bar{\nu} }{}^{ \bar{b} }
        - \delta_{ \bar{\mu} }{}^{ \bar{b} }
          \delta_{ \bar{\nu} }{}^{ \bar{a} }
       )
   = - (
          e_{ \bar{\mu} }{}^{ \bar{a} } e_{ \bar{\nu} }{}^{ \bar{b} }
        - e_{ \bar{\mu} }{}^{ \bar{b} } e_{ \bar{\nu} }{}^{ \bar{a} }     
       )
       / \ell^{2},
\\
\label{nonzero_curvature}
R_{ 5 \bar{\mu} }{}^{ \bar{a} \dot{5} }
   =   \mathfrak{a}^{\prime \prime} \delta_{ \bar{\mu} }{}^{ \bar{a} }
   =   e_{ \bar{\mu} }{}^{ \bar{a} }
       \left\{
            1 / \ell^{2}  
          - 2 [ \delta(x^{5}) - \delta(x^{5} - \pi \rho) ] / \ell
       \right\}, 
\end{gather}
\begin{gather}
\nonumber
R_{ \bar{\mu} \bar{\nu} }
   = - (
        \mathfrak{a} \mathfrak{a}^{\prime \prime}
        + 3 \mathfrak{a}^{\prime \, 2}
       )
       \eta_{ \bar{\mu} \bar{\nu} }
   = - \left\{
            4 / \ell^{2} 
          - 2 [ \delta(x^{5}) - \delta(x^{5} - \pi \rho) ] / \ell
       \right\}
       g_{ \bar{\mu} \bar{\nu} },
\\
R_{55} 
   = - 4 \mathfrak{a}^{-1} \mathfrak{a}^{\prime \prime}
   = - \left\{
            4 / \ell^{2}
          - 8 [ \delta(x^{5}) - \delta(x^{5} - \pi \rho) ] / \ell
       \right\},       
\end{gather}
\begin{equation}
R = - 8 \mathfrak{a}^{-1} \mathfrak{a}^{\prime \prime}
    - 12 \mathfrak{a}^{-2} \mathfrak{a}^{\prime \, 2}
  = - 20 / \ell^{2} 
    + 16 [ \delta(x^{5}) - \delta(x^{5} - \pi \rho) ] / \ell,
\end{equation}
and those related to (\ref{nonzero_curvature}) by
$ R_{\mu \nu}{}^{ab} = - R_{\nu \mu}{}^{ab} = - R_{\mu \nu}{}^{ba} $.
The prime symbol $ \prime $ denotes partial differentiation with respect
to $ x^{5} $.

\section*{Acknowledgements}

This work is supported by CNPq Edital Universal and FAPERJ.\\
F.T. acknowledges discussions with J. Zanelli about Reference 
\cite{CSADS5_OneBraneRand}.


\end{document}